# Thermal superscatterer: amplification of thermal scattering signatures for arbitrarily shaped thermal materials

Yichao Liu[1, #], Yawen Qi[1, #], Fei Sun[1, *], Jinyuan Shan[1], Hanchuan Chen[1], Yuying Hao[1], Hongmin Fei[1], Binzhao Cao[1], Xin Liu[1], Zhuanzhuan Huo[1]

*1 Key Lab of Advanced Transducers and Intelligent Control System, Ministry of Education and Shanxi Province, College of Physics and optoelectronic Engineering, Taiyuan University of Technology, Taiyuan, 030024, China*

[#] Yichao Liu and Yawen Qi contributed equally to this work.
[*] Corresponding author: sunfei@tyut.edu.cn;

**The concept of superscattering is extended to the thermal field through the design of a thermal superscatterer based on transformation thermodynamics. A small thermal scatterer of arbitrary shape and conductivity is encapsulated with an engineered negative-conductivity shell, creating a composite that mimics the scattering signature of a significantly larger scatterer. The amplified signature can match either a conformal larger scatterer (preserving conductivity) or a geometry-transformed one (modified conductivity). The implementation employs a positive-conductivity shell integrated with active thermal metasurfaces, demonstrated through three representative examples: super-insulating thermal scattering, super-conducting thermal scattering, and equivalent thermally transparent effects. Experimental validation shows the fabricated superscatterer amplifies the thermal scattering signature of a small insulated circular region by nine times, effectively mimicking the scattering signature of a circular region with ninefold radius. This approach enables thermal signature manipulation beyond physical size constraints, with potential applications in thermal superabsorbers/supersources, thermal camouflage, and energy management.**

## Introduction

Superscattering originates from a special optical illusion designed using transformation optics[1,2]. Unlike conventional scattering, superscattering refers to a phenomenon where the electromagnetic (EM) scattering cross-section of a structure, which consists of an original EM scatterer encapsulated by a metamaterial or metasurface shell, is far exceeds the geometrical cross-section of the entire structure[3,4]. Moreover, this enhanced scattering cross-section is entirely equivalent to the scattering cross-section of a transformed object, which is obtained by applying a spatial folding transformation or other composite coordinate transformations to the original EM scatterer[5]. This effect is typically achieved by encapsulating the original EM scatterer with a complementary medium pair consisting of a negative refractive index material and air, known as a superscatterer [1-5]. It should be noted that superscattering can also refer to cases where the scattering cross-section of a scatterer is significantly enhanced by surrounding it with plasmonic/dielectric shells[6] or gain metasurfaces[7], resulting in an overall scattering cross-section that surpasses the single-channel limit of subwavelength structures[8]. In contrast, superscatterers designed using transformation optics can produce an enhanced scattering cross-section that is identical to that of an expanded transformed object. Consequently, superscatterers designed using transformation optics can be extended to other optical illusions, such as superabsorbers[9] and invisible gateways[4,10].

In recent years, a similar concept has been extended to the thermal field as the thermal superscatterer, also known as the thermal magnifier[11] or amplifier[12]. Generally, thermal superscattering can be categorized as a specific type of thermal illusion[13]. This phenomenon is primarily achieved by encapsulating an original thermal scatterer of small size and thermal conductivity $\kappa_a$ (as shown in Fig. 1a) with an engineered shell. The resulting composite structure, formed through the integration of the original scatterer with an encapsulating engineered shell, is referred to as a thermal superscatterer. The encapsulating engineered shell with specially designed materials can be designed as a negative thermal conductivity shell (NTCS) in Fig. 1b or a positive thermal conductivity shell (PTCS) together with active thermal metasurfaces (ATMs) in Fig. 1c. As a result, the size of the region where the external heat flux is perturbed becomes significantly larger than that of the composite structure of the original small thermal scatterer and the engineered shell. This implies that the thermal scattering signature of the original small thermal scatterer, together with the encapsulating engineered shell, is greatly amplified. Moreover, the thermal scattering signature of the small thermal scatterer together with the engineered shell is entirely equivalent to that of an enlarged thermal scatterer with a scaled-up size. The enlarged thermal scatterer can either preserve the thermal conductivity while maintaining conformal geometry with the original small thermal scatterer, or undergo intentional modifications to both its shape and thermal conductivity (as shown in Fig. 1d). From the perspective of heat flux regulation and temperature field distribution, the temperature field outside the dashed lines in Figs. 1b and 1c is identical to that outside the enlarged thermal scatterer in Fig. 1d. It should be noted that in the thermal phenomena described in Figs. 1a-d, the regions outside the original thermal scatterer, the enlarged thermal scatterer and the thermal superscatterer are a uniform heat-conducting background medium with a thermal conductivity of $\kappa_b$.

With the advancement of transformation thermodynamics[14,15] and thermal metamaterials[16-20], many novel thermal illusion devices have been proposed, whose primary functions include reshaping the size and spatial distribution of thermal scatterers (thermal reshaper)[12,21], completely eliminating the thermal scattering signature of the scatterer (thermal cloaking)[22-26], rotating or concentrating heat flux (rotator and concentrator)[27-30], removing the original thermal signature of a scatterer while generating a new one[13,31], and achieving a Janus thermal function that depends on the direction of incident heat flux[32-34]. While pioneering theoretical investigations have laid important groundwork for understanding thermal superscattering through analytical solutions of Laplace's equation in cylindrical coordinates [11,12], several critical challenges remain to be addressed. First, existing analyses have been limited to canonical cylindrical/spherical geometries, leaving the more complex scattering behaviors of arbitrarily shaped structures unexplored. More significantly, the experimental realization of thermal superscatterers has proven challenging due to fundamental difficulties in realizing negative thermal conductivity, resulting in no experimental demonstrations reported. To further expand the range of achievable thermal conductivity, our recent research demonstrates that negative thermal conductivity can be equivalently realized by ATMs[35], which has already been successfully used to realize a long-focus thermal lens[36]. In this study, we design ATMs with precisely tailored spatial distributions and thermal output powers (as illustrated in Fig. 1c) to effectively replicate the functionality of negative thermal conductivity (as depicted in Fig. 1b) and subsequently demonstrate a thermal superscatterer experimentally.

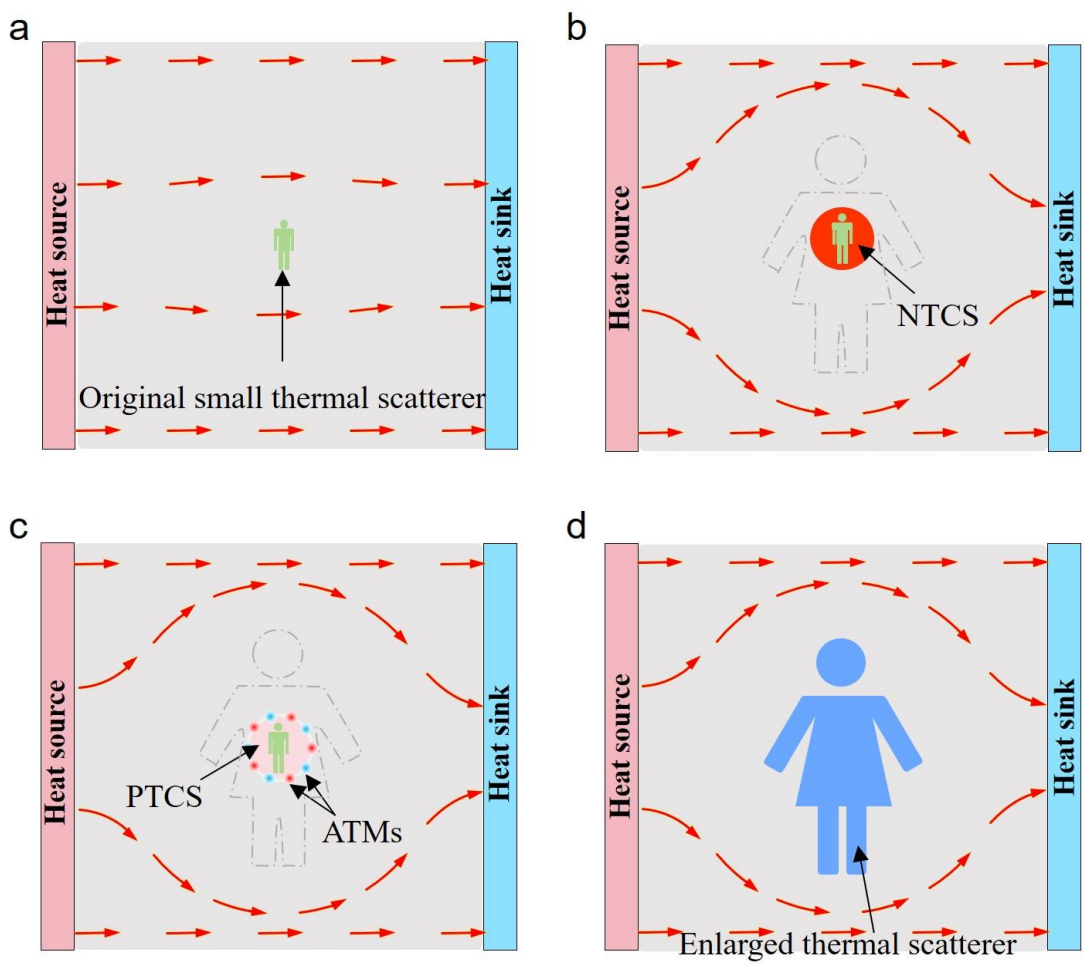

**Fig. 1 | Schematic diagram of the thermal superscatterer.** The red arrows indicate the direction of the heat flux, and the gray dashed lines in (b) and (c) serve as an equivalent thermal scattering cross-section for comparison. (a) A small thermal object (indicated by a small green man) with thermal conductivity $\kappa_1$ in a uniform background thermally conducting medium with thermal conductivity $\kappa_b$ functions as the original

thermal scatterer. (b) The original thermal scatterer is covered by NTCS (the red shell) in the same background thermally conducting medium, resulting in thermal superscattering. (c) The original thermal scatterer is surrounded by ATMs in the same background thermally conducting medium, also resulting in thermal superscattering. (d) An enlarged thermal scatterer (indicated by a large blue woman) with thermal conductivity $\kappa_a$ in the same background thermally conducting medium can produce the same thermal scattering signature (i.e., the same temperature distribution and heat flux distribution) as the region outside the dashed lines in (b) and (c). More details are given in the Supplementary Movie 1.

## Results

### General design method

Firstly, we design a NTCS capable of achieving thermal superscattering in Fig. 1b by transformation thermodynamics. The design is based on a steady-state system described in cylindrical coordinates ($\rho$, $\theta$, $z$) under a two-dimensional (2D) configuration. Notably, the implementation follows an inverse approach, which reverses the design sequence from panels (a) to (d) in Fig. 1. Specifically, the enlarged thermal scatterer in Fig. 1d is modeled as a thermal object characterized by a thermal conductivity of $\kappa_a$ and a boundary defined by $\rho_3(\theta')$. It is centrally embedded within the background material, which has a thermal conductivity of $\kappa_b$, in the reference space, as illustrated in Fig. 2a. The original small thermal scatterer and the NTCS in Fig. 1b are modeled as the thermal objects in the physical space illustrated in Fig. 2b. The thermal conductivity of the original small thermal scatterer and the NTCS are $\kappa_1$ and $\kappa_2$, respectively, while their outer boundaries are represented by $\rho_1(\theta)$ and $\rho_2(\theta)$, respectively. Now, we will demonstrate how to establish the correspondence between the reference space in Fig. 2a and the physical space in Fig. 2b through a continuous composite coordinate transformation.

To ensure that the original small thermal scatterer encapsulated with the NTCS in the physical space (Fig. 2b) can produce an identical thermal scattering signature to the enlarged thermal scatterer in the reference space (Fig. 2a), a composite coordinate transformation can be employed. This transformation involves folding the boundary $\rho_3(\theta')$ inward to $\rho_1(\theta')$ with $\rho_2(\theta')$ as the fixed boundary, while simultaneously compressing the region enclosed by $\rho_3(\theta')$ into the interior of $\rho_1(\theta')$. Note the symbols with and without primes denote the coordinates in the reference space and physical space, respectively, and the angular coordinates remain the same during the transformation, i.e., $\theta = \theta'$. The detailed coordinate transformation formula can be written as,

$$\begin{cases} \rho = \dfrac{\rho_1(\theta')}{\rho_3(\theta')}\rho' & \text{for } \rho \leq \rho_1(\theta) \\ \rho = \dfrac{\rho_2(\theta')^2}{\rho'} & \text{for } \rho_1(\theta) < \rho \leq \rho_2(\theta) \\ \rho = \rho' & \text{for } \rho > \rho_2(\theta) \end{cases}, \quad \theta = \theta', \text{ and } z = z'. \quad (1)$$

The transformation continuity condition requires the boundary parameters to satisfy the geometric relation: $\rho_1\rho_3=\rho_2^2$. The thermal conductivity for the small thermal scatterer $\kappa_1$ within region $\rho < \rho_1(\theta)$, the NTCS $\kappa_2$ within region $\rho_1(\theta) < \rho < \rho_2(\theta)$, and the background $\kappa_3$ within region $\rho > \rho_2(\theta)$, can be calculated using transformation thermodynamics[37] (details can be found in the **Method section**),

$$\kappa_1 = \begin{pmatrix} 1+\left[\dfrac{\partial(\rho_1/\rho_3)}{\partial\theta}\right]^2\left(\dfrac{\rho_3}{\rho_1}\right)^2 & \dfrac{\partial(\rho_1/\rho_3)}{\partial\theta}\dfrac{\rho_3}{\rho_1} \\ \dfrac{\partial(\rho_1/\rho_3)}{\partial\theta}\dfrac{\rho_3}{\rho_1} & 1 \end{pmatrix}\kappa_a, \quad \text{for } \rho<\rho_1(\theta), \quad (2)$$

$$\kappa_2 = -\begin{pmatrix} 1+\dfrac{4}{\rho_2^2}\left(\dfrac{\partial\rho_2}{\partial\theta}\right)^2 & \dfrac{2}{\rho_2}\dfrac{\partial\rho_2}{\partial\theta} \\ \dfrac{2}{\rho_2}\dfrac{\partial\rho_2}{\partial\theta} & 1 \end{pmatrix}\kappa_b, \quad \text{for } \rho_1(\theta)<\rho<\rho_2(\theta), \quad (3)$$

$$\kappa_3 = \kappa_b, \quad \text{for } \rho > \rho_2(\theta). \quad (4)$$

The thermal conductivity of both the NTCS described in Eq. (2) and the small scatterer described in Eq. (3) are typically angular-dependent tensors. The background thermally conducting medium outside the equivalent thermal scattering boundary (outside the dashed boxes in Fig. 2b) after encapsulating the NTCS is identical to the background thermally conducting medium outside the region of the large-sized scatterer with an outer boundary at $\rho_3$. This is a result of the identity transformation applied beyond $\rho_3$. Eq. (4) further demonstrates that the thermally conducting material outside the boundary $\rho_2$ in Fig. 2b is identical to that outside the boundary $\rho_3$ in Fig. 2a, which means the same thermal scattering signature can be obtained with a decreased overall geometrical size.

Second, as illustrated in Fig. 1c, the NTCS is equivalently replaced by a PTCS combined with a boundary heat source $q_s$ (i.e., boundary heat flux intensity). To maintain both the original temperature distribution and energy conservation, the thermal conductivity of the PTCS must satisfy $\kappa_{PTCS} = -\kappa_2$, while the boundary heat source is determined by $q_s = -2q_n$[35]. Here, $q_n$ represents the normal component of heat flux at the NTCS boundary, adhering to the sign convention where inward-directed fluxes toward the NTCS are defined as positive.

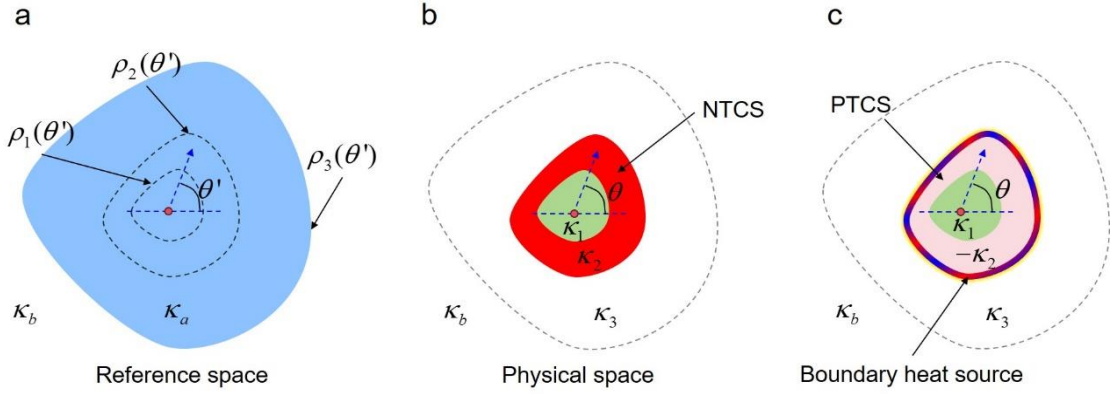

**Fig. 2 | Design of the NTCS and the PTCS with a boundary heat source.** (a) A large thermal scatterer (colored blue) with thermal conductivity $\kappa_a$ in the reference space, which corresponds to the enlarged thermal scatterer in Fig. 1d. (b) A small thermal scatterer (colored green) with thermal conductivity $\kappa_1$ covered by a NTCS (colored red) in the physical space, which are transformed from the large thermal scatterer by the coordinate transformation in Eq. (1). (c) The NTCS is replaced by a PTCS (colored pink) with a boundary heat source. All the three cases have the same thermal scattering signature outside the boundary $\rho_3(\theta)$. The white regions are the same background thermally conducting medium with thermal conductivity $\kappa_b$.

**Two special cases with simplified material parameters**

Considering the complexity of the material parameters of the small scatterer and the engineered shell that display inhomogeneity and anisotropy in the absence of specific selection of the three boundaries, we focus on examining two special cases to achieve further simplification of these material parameters in Eqs. (2) and (3). The first case involves the conformal enlarging of the geometric region of the original small thermal scatterer with arbitrary thermal conductivity while keeping the thermal conductivity unchanged. In this case, the original small thermal scatterer and the enlarged thermal scatterer are homothetic. Specifically, it includes three examples: super-insulating thermal scattering, super-conducting thermal scattering, and equivalent thermally transparent effect. The second case involves the simultaneously enlarging and reshaping the thermal scattering signature of the original small thermal scatterer with intentionally modified thermal conductivity. Specifically, if the original small thermal scatterer is made of adiabatic material, the enlarged thermal scatterer can also maintain adiabatic properties under coordinate transformation.

In the first case, the shapes of the original thermal scatterer (indicated by the boundary $\rho_1(\theta)$), the engineered shell (indicated by the boundary $\rho_2(\theta)$), and the enlarged thermal scatterer (indicated by the boundary $\rho_3(\theta)$) are conformal. Their conformal nature implies that the ratio $\rho_1(\theta)/\rho_3(\theta)$ is independent of the angular coordinate $\theta$. Consequently, the partial derivative $\partial(\rho_1(\theta)/\rho_3(\theta))/\partial\theta = 0$, which leads to $\kappa_1 = \kappa_a$ through Eq. (2). This implies that if the shapes of the small thermal scatterer and the enlarged thermal scatterer are conformal, the thermal conductivity $\kappa_a$ of the enlarged thermal scatterer in the reference space of Fig. 2a is identical to the thermal conductivity $\kappa_1$ of the small thermal scatterer in the physical space of Figs. 2b and 2c. In other words, the thermal superscatterer only magnifies the geometrical size of the thermal scattering signature without altering the material or shape of the scatterer. Under this condition, we simulate three cases where the small thermal scatterer is filled with an adiabatic material ($\kappa_a=0$), a highly conductive material ($\kappa_a = 1000\kappa_b$), and a background thermally conductive material ($\kappa_a = \kappa_b$), respectively (see Fig. 3).

Fig. 3a shows the simulated temperature distribution when the large thermal scatterer with $\kappa_a = 0$ is exposed to a uniform thermal flux. Fig. 3b illustrates the temperature distribution when replacing the large thermal scatterer with a smaller thermal scatterer of the same thermal conductivity, and a NTCS whose thermal conductivity follows Eq. (3). The resulting temperature distribution closely resembles that of the large thermal scatterer. In Fig. 3c, the NTCS is replaced by a corresponding PTCS, as well as a boundary heat source. Note only the boundary heat source for boundary $\rho_2(\theta)$ needs to be considered when $\kappa_a = 0$, as the normal boundary heat flux along boundary $\rho_1(\theta)$ is zero. In contrast, when $\kappa_a \neq 0$, both boundaries must be taken into account. In this example, with the aid of thermal super-scatterers shown in Figs. 3b-c, adiabatic materials occupying a smaller area encapsulated with the engineered shells can achieve the same thermal shielding effect as those covering a larger area for the external background region, thereby realizing the super-insulating thermal scattering effect (or super-thermal shielding effect). In this case, the isothermal line distribution in the external region $\rho > \rho_3$ shown in Figs. 3b-3c is identical to that in Fig. 3a where heat flux cannot penetrate the same area. This creates an illusion that external heat flow cannot enter the region $\rho < \rho_3$, while in reality the structure remains capable of sensing external heat flow - achieving the "thermally inaccessible yet thermally perceptive" phenomenon. Therefore, this configuration can be used to create virtual thermal insulation boundaries. Moreover, in this example, the geometric dimensions of the engineered shell $\rho_2$ are smaller than those of the virtual adiabatic boundary $\rho_3$. This creates a transitional zone between the engineered shell and the virtual insulation boundary that maintains the same thermal conductivity as the background material. This region appears thermally inaccessible from the external perspective while actually permitting heat flux penetration -precisely embodying the essential meaning of "super" in the term super-scattering phenomenon.

In Fig. 3d, the thermal conductivity of the enlarged thermal scatterer is chosen as a high value, i.e., $\kappa_a = 1000\kappa_b$. Two cases demonstrate that a small thermal scatterer, enveloped by a NTCS in Fig. 3e and a PTCS together with a suitably designed boundary heat source in Fig. 3f, respectively, can both achieve a temperature distribution similar to that of the large thermal scatterer with the same high thermal conductivity in Fig. 3d. In this example, with the aid of thermal super-scatterers shown in Figs. 3e-f, materials with high thermal conductivity occupying a smaller area encapsulated with the engineered shells can achieve the same good heat conduction effect as those covering a larger area for the external background region, thereby realizing the super-conducting thermal scattering effect. This means that high thermal conductivity materials occupying a smaller area ($\rho < \rho_1$), when encapsulated with the engineered shells

($\rho_1<\rho<\rho_2$), can be used to achieve the same heat flow attraction and convergence effects as those of materials with the same thermal conductivity occupying a larger area ($\rho<\rho_3$). Notably, the geometric dimension of the engineered shell $\rho_2$ is intentionally designed with geometric dimension smaller than those of the enlarged high-conductivity thermal scatterer $\rho_3$. The intervening space between them ($\rho_2<\rho<\rho_3$) is occupied by a low thermal conductivity background material $\kappa_b$, which is far smaller than the thermal conductivity of enlarged thermal scatterer $\kappa_a = 1000\kappa_b$. Therefore, this configuration effectively mimics a virtual high-conductivity boundary, enabling applications such as efficient heat energy collectors (effectively increasing the heat concentration cross-sectional area of the thermally conductive material), integration of thermal sinks as super-thermal absorbers, or embedding heat generators as super-thermal sources.

More interestingly, if the large thermal scatterer vanishes ($\kappa_a = \kappa_b$) in Fig. 3g, we can still design a NTCS in Fig. 3h and a PTCS together with a suitably designed boundary heat source in Fig. 3i, respectively to achieve an equivalent thermally transparent effect. In this case, although the original small thermal scatter, when encapsulated with the engineered shells in Figs. 3h-i, do not affect the distribution of the temperature field and heat flux in the external region $\rho>\rho_3$. Notably, in this configuration, the annular region between the engineered functional shell and the enlarged thermal scatterer, i.e., $\rho_2<\rho<\rho_3$, also preserves the original thermal state, maintaining both the temperature field and heat flux distribution. However, through elaborate design of the boundary heat sources along boundaries $\rho_2$ and $\rho_3$, we can achieve redistribution of the temperature field and heat flux inside boundary $\rho_2(\theta)$. That is, it is thermally transparent to the external thermal background, the gradient distribution of the internal temperature field can be adjusted, thus enabling local thermal management without external disturbance. Note changing the thermal conductivity of the large/small scatterer will not influence the material parameter of the NTCS and PTCS, which is only determined by the thermal conductivity of the background and the profile of the boundary $\rho_2(\theta)$.

Although the first three case studies adopt a specific geometric configuration, the proposed method demonstrates universal applicability to thermal superscatterers of arbitrary shapes. As detailed in Supplementary Notes Section 1 (Thermal superscatterers with different geometries), we have successfully designed superscatterers with alternative geometries - including triangular and rectangular configurations - through systematic numerical simulations and theoretical analyses.

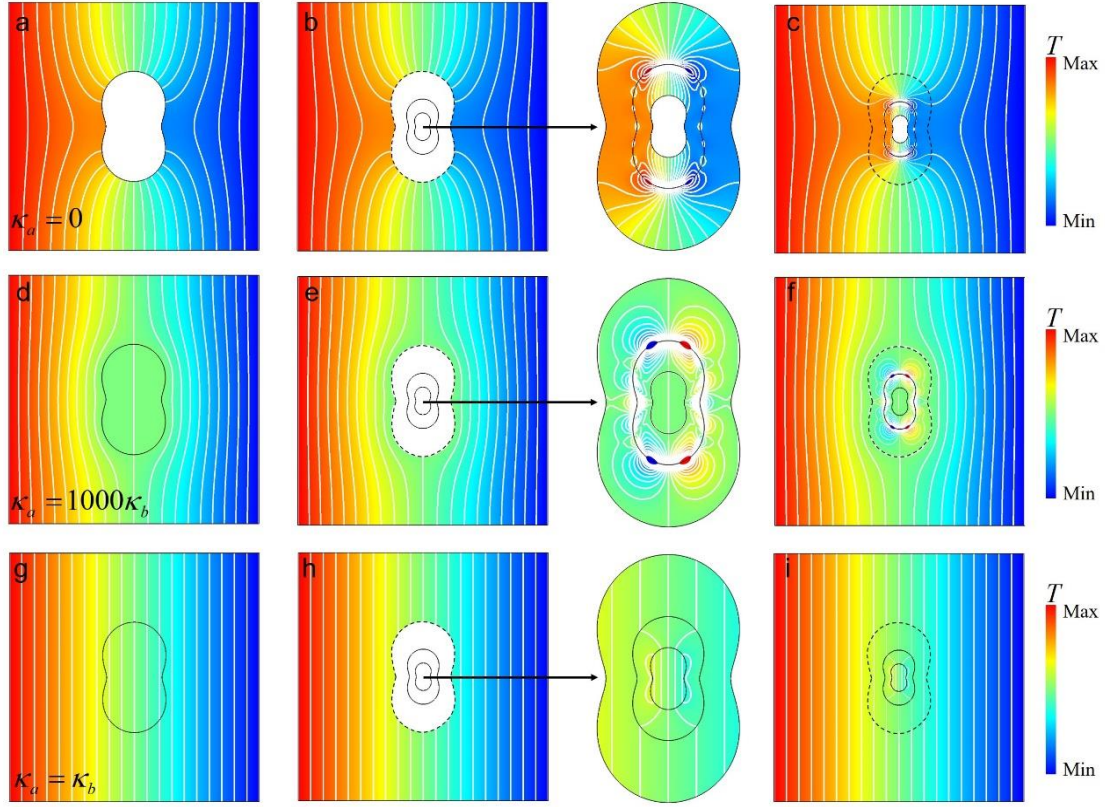

**Fig. 3 | Simulation results for conformal shapes.** Simulated temperature distributions for peanut-shaped large scatterers with different thermal conductivity: (a) $\kappa_a = 0$, (d) $\kappa_a = 1000\kappa_b$, and (g) $\kappa_a = \kappa_b$. In panels (b), (e), and (h), the small scatterers exhibit the same thermal conductivity as the large thermal scatterers and are covered by the same NTCS. For enhanced clarity, the regions within the boundary $\rho_3(\theta)$ are enlarged and presented as insets. In (c), (f), and (i), PTCS and boundary heat sources are used to achieve the corresponding thermal superscattering effect, which are the same as that in (b), (e), and (h).

The second case corresponds to $\rho_2(\theta)$ = const, indicating that the engineered shell now possesses a circular outer boundary with a radius of $R_2$. In this case, the thermal conductivity of the NTCS region can be simplified to be the negative of the background material, i.e., $\kappa_2 = -\kappa_b$, which can be derived from Eq. (3) by setting the partial derivative $\partial \rho_2(\theta)/\partial \theta = 0$. Then, the modified thermal conductivity of enlarged thermal scatterers $\kappa_a$ can be designed through the transformation in Eq. (2). Such modification can be intentionally achieved by designing the ratio $\rho_1(\theta)/\rho_3(\theta)$ and the thermal conductivity of smaller thermal scatterers $\kappa_1$. As a special case, when the small thermal scatterers are adiabatic materials (i.e., $\kappa_1=0$), Eq. (2) reveals that the corresponding enlarged thermal scatterers also exhibit adiabatic characteristics (with unchanged material thermal conductivity $\kappa_a=0$). Meanwhile, the small thermal scatterers often exhibit geometrical shapes that differ from those of the enlarged thermal scatterers except for a circular shape. For example, the small four-petal-shaped scatterer in Figs. 4b-c can generate identical thermal scattering signatures to the enlarged square in Fig. 4a, while the small three-petal-shaped scatterer in Figs. 4e-4f produces identical thermal scattering features to the enlarged triangle in Figs. 4d.

Figs. 4a, 4d, and 4g present the simulation results for three large thermal scatterers with different shapes, each having a thermal conductivity of $\kappa_a = 0$. When these large thermal scatterers in Figs. 4a, 4d, and 4g are replaced by the small thermal scatters combined with engineered shells, the corresponding temperature distributions associated with the NTCS are illustrated in Figs. 4b, 4e, and 4h, whereas the corresponding temperature distributions with boundary heat sources are shown in Figs. 4c, 4f, and 4i. The close resemblance to the large scatterers depicted in Figs. 4a, 4d, and 4g demonstrates that the thermal superscattering effect can be achieved using only a boundary heat source at $\rho_2(\theta) = $ const, where the PTCS is replaced by the background materials (i.e., $\kappa_{PTCS} = \kappa_b$).

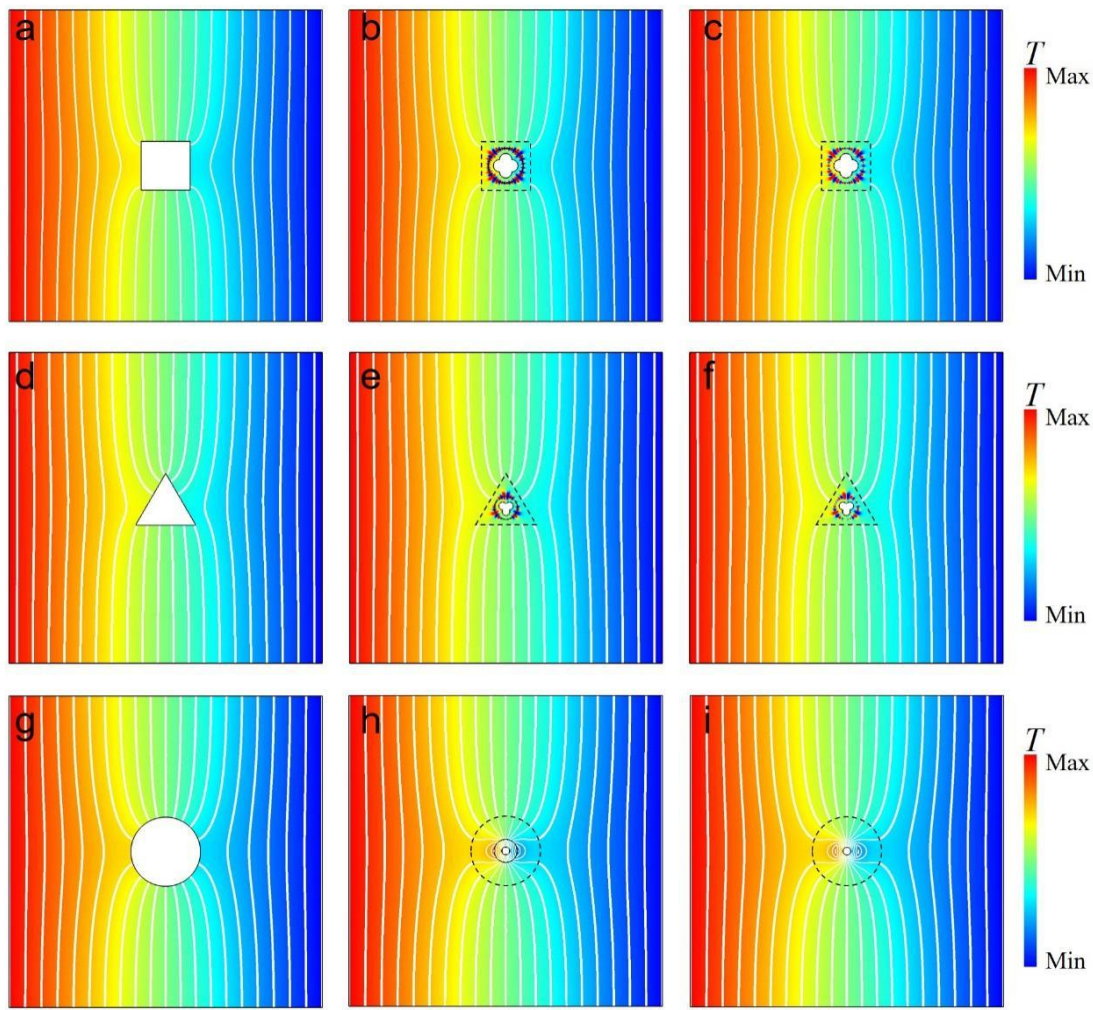

**Fig. 4 | Simulation results for boundary $\rho_2(\theta) = $ const.** (a), (d), and (g) are the simulated temperature distributions for large scatterers with different shapes. (b), (e), and (h) are the corresponding superscatterers consist of small scatterers and NTCSs. (c), (f), and (i) are the corresponding superscatterers consist of small scatterers and circular boundary heat sources. For enhanced clarity of the temperature distribution within the superscatterers, the isotherms inside the boundary $\rho_3(\theta)$ are removed in (b), (c), (e), and

(f). Note that the small thermal scatterers are designed as adiabatic materials, which leads to the corresponding enlarged thermal scatterers exhibiting adiabatic characteristics according to Eq. (2).

**Design of the ATMs array**

To experimentally realize the superscattering effect, an array of ATMs should be designed to equivalently achieve the boundary heat source. For ease of fabrication and processing, this study uses the small circular thermal scatterer filled with adiabatic materials in Fig. 4i as an example to demonstrate the design methodology for replacing boundary heat sources with ATMs and to validate the approach through subsequent experimental verification. Here, the boundary heat source at the boundary $\rho_2(\theta) = R_2$ in Fig. 5a is discretized into $M$ curve segments (with equal angular spacing $\Delta\theta=2\pi/M$ and equal arc length $\Delta s=\Delta\theta R_2$), where an ATM is placed at the midpoint of each segment, as shown in Fig. 5b. The heat power of each ATM is obtained by integrating the boundary heat flux density $q_s$ over the corresponding curved segment[35]:

$$Q_m = \int_{C_m} q_s \mathrm{d}s, \tag{5}$$

where $C_m$ represents the $m$-th curved segment on the circular arc of radius $R_2$, and $\mathrm{d}s$ is the arc length differential element on $C_m$. In our previous studies, it is found that a larger $M$ enables a more accurate approximation of the effects generated by a continuous boundary heat source using discrete ATM[35]. However, considering the practical constraints of experimental fabrication, $M$ should not be chosen excessively large. Balancing these two considerations, $M = 10$ is selected for subsequent numerical and experimental studies. Further details regarding the influence of the discretization number $M$ on superscatterer performance are provided in Supplementary Notes Section 2 (Impact of Discretization Number $M$ on Thermal Superscatterer Performance).

Before conducting experiments, the effectiveness of the method by replacing the boundary heat sources in Fig. 5a with ATMs in Fig. 5b is validated through numerical simulations. For the configurations in Figs. 5a and 5b under identical thermal flux incidence from left to right (with boundary conditions specified in the numerical settings), the simulated temperature fields and isotherms are shown in Figs. 5c and 5d, respectively. The simulated temperature distribution and isotherms outside the boundary $R_2$ for the case using discrete ATMs in Fig. 5d are fully consistent with those for the case using continuous boundary heat sources in Fig. 5c. The simulated results verify that replacing the boundary heat sources in Fig. 5a by discretizing the boundary into $M = 10$ equal segments and allocating the total heat flux of each segment to an ATM positioned at its midpoint in Fig. 5b can effectively replicate the effects of continuous boundary heat sources, thereby enabling the super-insulating thermal scattering effect in Fig. 4i. Therefore, the ATM configuration shown in Fig. 5b will be adopted for subsequent experimental validation on super-insulating thermal scattering effect.

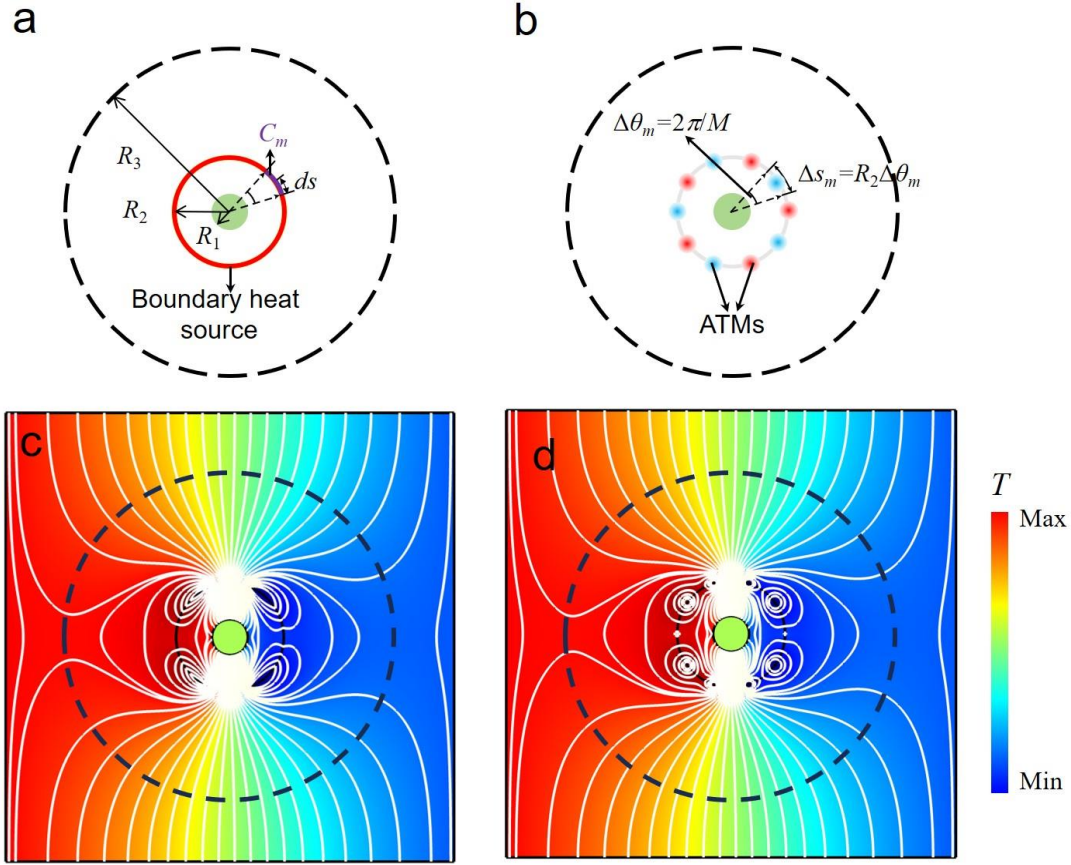

**Fig. 5 | Design diagram of the ATMs array.** (a) Structural diagram of the thermal superscatterer with boundary heat source. The red line represents the boundary heat source. (b) Structural diagram of the thermal superscatterer with ATMs. Red/blue dots represent ATMs. In (a) and (b), all materials except for the central adiabatic region have the same thermal conductivity as the background material, which is represented by the white areas. The spacing angle between adjacent ATMs is $\Delta\theta = 36°$. And $\Delta s$ represent the corresponding arc length for each ATMs. Since the small thermal scatterer, engineered shell, and enlarged thermal scatterer all exhibit circular geometries in this case, their boundary profiles reduce to constant radial functions independent of $\theta$: $\rho_1(\theta) = R_1$, $\rho_2(\theta) = R_2$ $\rho_3(\theta) = R_3$. (c) Simulated result of a thermal superscatterer with a continuous boundary heat source. (d) Simulated temperature distribution and isotherms for a thermal superscatterer with 10 ATMs. In (c) and (d), the white lines represent the isotherms. The black dashed small and big circles represent the boundary where heat source or ATMs are located $\rho_2(\theta) = R_2$ and the virtual enlarged boundary $\rho_3(\theta) = R_3$, respectively.

## Experimental verification

By utilizing the design shown in Fig. 5b, the sample is fabricated in Figs. 6a-6b and the experimental measurement system is established in Fig. 6c. The sample preparation and assembly process are given in the sample preparation section, shown in Fig. 6a. Here, the upper copper sheet serves as the background thermal conduction medium, while the bottom copper plate acts as a heat dissipation sink. Expanded Polyethylene (EPE) foam sheet and expanded polystyrene (EPS) foam are used for supporting and simulating thermal insulation, respectively. 10 semiconductor cooling plates are utilized as the ATMs, which can generate thermal output power as designed by Eq. (5) using tunable loading current [38] and are evenly fixed on the circle $R_2 = 30$ mm. The assembled sample is shown in Fig. 6b, positioned within the detection region of the experimental measurement system illustrated in Fig. 6c. A mechanically rigid material (polylactic acid fiber with a thermal conductivity of 0.13 W·m$^{-1}$·K$^{-1}$ and a height of 20 cm) serves as Support B to ensure alignment between the upper surface of the ATMs and the upper surfaces of supports A1 and A2 (polylactic acid fiber with a thermal conductivity of 0.13 W·m$^{-1}$·K$^{-1}$). The function of Supports A1 and A2 on the sidewalls of the two water baths is to allow the two ends of the copper sheet to be smoothly immersed in the water along the rounded edges of the supports. This configuration also provides sufficient space beneath the copper plate for heat convection, enabling the copper plate's temperature to stabilize near ambient temperature. The experiment is conducted in an enclosed room with a constant temperature $T_{room} = 300.65$ K.

An electric constant-temperature water bath (labeled as 'hot' colored red) serves as the heat source, while a large insulated water bath (labeled as 'cold' colored blue) acts as the heat sink. Throughout the experiment, the temperatures of the heat source and heat sink remain nearly constant at 320 K and 287 K, respectively. This ensures that the left and right boundaries of the detection region maintain stable constant temperatures of 307 K and 297 K. The temperature difference between the heat source and heat sink induces a heat flux in the detection region, which aligns with the simulation results presented in Figs. 3-5. To minimize the influence of natural convection, the experimental detection region is enclosed with EPS foam boards. Subsequently, experimental measurements are conducted on the assembled sample, which is positioned within the detection region and observed by the infrared camera (FOTRIC 288). Based on the required heat power $Q_m$ calculated from Eq. (5) for each ATM, the corresponding current for each port of DC power supply is set according to the calculated value[35,38]. After energizing the ATMs, a 30-second waiting period is required until the temperature field distribution observed by the infrared camera stabilizes, i.e., the system reaches a steady state. Then, the temperature distribution of the detection region is recorded by the infrared camera. Due to the neglect of factors in the initial current calculations, such as environmental convection disturbances and sample preparation errors, appropriate adjustments should be made to the applied current of DC power supply. To address this discrepancy, the proposed methodology involves comparing the temperature distribution obtained from the initial experimental measurements with the simulation results in Fig. 5d. Through systematic analysis of isotherm distribution disparities, the applied currents for relevant ATMs are fine-tuned, ensuring close alignment between the experimental and simulated isotherm distributions. For more details regarding experimental measurement, please refer to Supplementary Notes Section 4 (Details of Experimental Measurements) and Supplementary Movie 2.

The measured temperature distribution of the original small thermally insulated circular region with radius $R_1$ = 10 mm, when encapsulated with precisely designed 10 ATMs (functioning as the engineered shell) in Fig. 6d, shows consistency with the corresponding simulation result in Fig. 5d. For comparison, reference measurements are also conducted for three additional cases: a copper sheet with a large circular thermally insulated region ($R_3$ = 90 mm) in Fig. 6e, a copper sheet with a small circular thermally insulated region ($R_1$ = 10 mm) in Fig. 6f, and an intact copper sheet without any thermally insulated region in Fig. 6g.

Compared with the case without any thermal scatter in Fig. 6g, the measured result in Fig. 6f demonstrate that the small thermal scatterer exhibits negligible impact on both the temperature distribution and the isotherms. However, when the small circular thermally insulated region with radius $R_1$ = 10 mm is scaled up by a factor of nine, the enlarged thermal scatterer with radius $R_3$ = 90 mm significantly alters both the temperature distribution and the isotherms in Fig. 6e. Notably, the temperature and isotherm distributions in Fig. 6d are nearly consistent with those in Fig. 6e. This indicates that the thermal scattering signature of the larger thermally insulated circular region in Fig. 6e can be effectively replicated by the original small thermally insulated circular region encapsulated with precisely engineered ATMs in Fig. 6d. Therefore, the experimental results conclusively demonstrate that a small thermal insulated scatterer enclosed by precisely engineered ATMs can successfully emulate a larger thermal insulated scatterer, thereby experimentally validating the super-insulating thermal scattering effect. Notably, the super-insulating thermal scattering effect achieved here enables the creation of "virtual thermal insulation boundaries," where external heat flux perceives an enlarged insulated region, while the actual structure retains compact dimensions. Such capability opens avenues for applications in thermal camouflage, where localized heat management can be achieved without altering macroscopic material layouts.

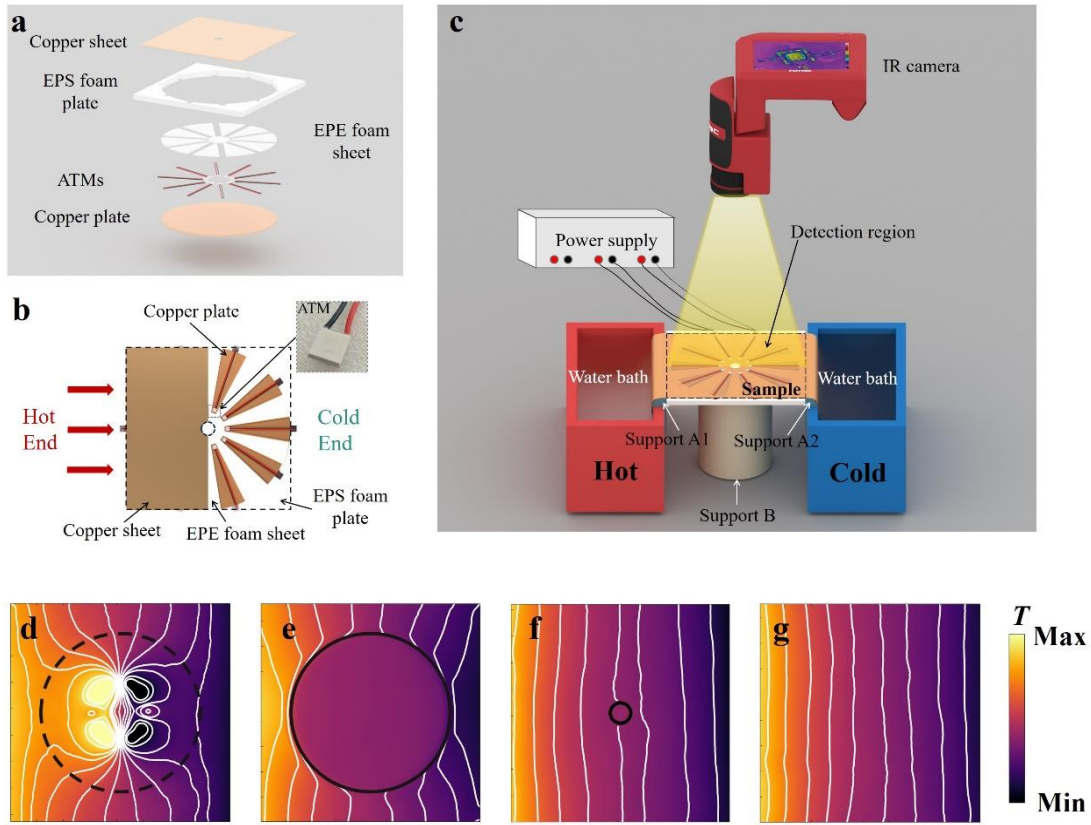

**Fig. 6 | Experimental setup and results.** (a) Assembly details of the sample. (b) Schematic diagram of the designed thermal superscatterer with 10 ATMs (top view). The dashed line represents the complete shape of the copper sheet in the detection region. 10 ATMs are evenly distributed along a circle with radius of $R_2$ = 30 mm. (c) Schematic diagram of the experimental setup. The detection region is indicated by the dashed lines. The water baths on both sides provide constant temperature heat sources for the experiment. (d)-(g) each depict the temperature distribution and isotherms as measured in the detection region. (d) The temperature distribution and isotherms of the thermal superscatterer. Dashed circle is plotted for comparison. (e) There is a 90 mm-radius circular hole at the center of copper sheet. (f) There is a 10 mm-radius circular hole at the center of the copper sheet. (g) The detection region is entirely made up of copper sheet. Temperature distributions and isotherms depicted in (e)-(g) all serve as references for the effect of thermal superscatterer.

## Discussion

In summary, the concept of superscattering has successfully been extended to the thermal field by designing and experimentally validating a thermal superscatterer based on transformation thermodynamics and ATMs. The method for the realization of thermal superscattering effect involves no negative thermal conductivity materials, thus making its realization feasible. The design framework supports three distinct thermal illusions: super-insulating scattering (mimicking enlarged adiabatic regions and achieving the "thermally inaccessible yet thermally perceptive" by virtual thermal insulation boundaries), super-conducting scattering (emulating large high-conductivity domains by virtual high-conductivity boundaries), and equivalent thermal transparency (masking internal thermal perturbations). The experimental results validate that a small thermally insulated circular region encapsulated with an array of 10 precisely engineered ATMs can effectively magnifies the thermal scattering cross-section of an insulated circular region by a factor of 9. The method in this study can be extended to other applications, such as super thermal insulation, thermal invisibility gateways, thermal superabsorbers and thermal supersources.

## Method

### Derivation of the thermal conductivity

The derivation is performed in the cylindrical coordinate system, where the metric tensor in the reference space and physical space can be expressed as $g' = diag\{1, \rho'^2, 1\}$, $\gamma = diag\{1, \rho^2, 1\}$, respectively. The thermal conductivity in the cylindrical coordinate system can be expressed as,

$$\kappa_{cyl.} = \frac{\sqrt{\det g'}}{\sqrt{\det \gamma}} \frac{Jg^{-1}J^T}{\det J} \kappa_{ref}, \qquad (6)$$

where $J$ is the Jacobian matrix of the transformation and $\kappa_{ref}$ is the thermal conductivity in the reference space. When using the normalized coordinate basis, the thermal conductivity can be written as,

$$\kappa = \Lambda \kappa_{cyl.} \Lambda^T, \qquad (7)$$

where $\Lambda = diag\{1, \rho, 1\}$. Eq. (7) can be used to derive thermal conductivity for different regions with different transformations in Eq. (1).

### Numerical settings

The numerical simulations are all conducted by commercial software COMSOL Multiphysics 5.6 with the license number 9406999 (https://www.comsol.com), which is based on the finite element method. The 2D solid heat transfer module with steady-state solver is selected to simulate the temperature field distributions in all the simulations. The free tetrahedral meshing with automatically generated mesh grid is used.

For all the simulations in this study, the left boundary is set to a constant high temperature of 307.65 K, the right boundary is set to a constant low temperature of 297.65 K, and the background thermal conductivity is set to 400 W·m$^{-1}$·K$^{-1}$. In Fig. 3, $\rho_3(\theta)/\rho_2(\theta) = \rho_2(\theta)/\rho_1(\theta) = 2$. The side length of the square $l_1$ in Fig. 4a-4c, the side length of the triangle $l_2$ in Fig. 4d-4f, and the radius of the outer circle $R_3$ in Fig. 4g-4i, has the following relationship with the circle $\rho_2(\theta) = R_2$, i.e., $l_1 = 2\sqrt{2}R_2$, $l_2 = 3\sqrt{3}R_2$, $R_3 = 3R_2$.

**Sample preparations**

Firstly, we prepare a round copper plate with a radius of 125 mm and a thickness of 3 mm for the heat dissipation of ATMs. Then we mark 10 points on the copper plate to indicate the positions of the ATMs. These 10 points are arranged in circle $\rho_2(\theta) = R_2$ with the equal spacing angle $\Delta\theta = 36°$ as shown in Fig. 5b. The semiconductor cooling plate (TECooler Technology, model HT009022, size 9.8 mm × 9.8 mm × 2.59 mm) is selected to be utilized as ATMs. The 10 ATMs are positioned at the designated locations on the copper plate. Then, the remaining space between the ATMs is filled with Expanded Polyethylene (EPE) foam sheet that has a thickness of 2 mm and a thermal conductivity of 0.05 $W·m^{-1}·K^{-1}$. And we use an expanded polystyrene (EPS) foam board with a thickness of 10 mm and a thermal conductivity of 0.035 $W·m^{-1}·K^{-1}$ to encase the lateral air region surrounding the copper plate in the detection region. Both EPE foam sheet and EPS foam board serve to reduce heat exchange and provide support for the copper sheet, with their upper surfaces align with the upper surface of the ATMs. A copper sheet with size of 550 mm × 250 mm × 0.1 mm, and featuring a circular hole with a radius of $R_1 = 10$ mm is positioned above the ATMs. To reduce thermal resistance, thermal conductive silicone grease is used to fill the air gaps between the ATMs and both the copper plate beneath and the copper sheet above it. To ensure a uniform high surface emissivity, we apply black body paint to the upper surface of the copper sheet. For additional details regarding sample fabrication and experimental setup, please refer to Supplementary Notes Section 3 (Sample Preparation and Pre-Experiment Preparation).

**Data availability**

The data supporting the findings of this study are available from the corresponding authors upon reasonable request.

**Acknowledgments**
This work is supported by the National Natural Science Foundation of China (Nos. 12374277, 12274317, 61971300, 61905208 and 11604292) and Natural Science Foundation of Shanxi Province (202303021211054).
**Author contributions**
Y. L. and Y. Q. performed the simulations. Y. Q. and Z. H. performed the measurements. Y. Q., H. C., and J. S. prepared the sample. Y. L. and F. S. proposed the idea. Y. Q., Y. L., and F. S. prepared the manuscript, Y. L., Y. Q., F. S., J. S., H. C., Y. H., H. F., B. C., X. L., and Z. H. contributed to the discussion.
**Competing interests**
The authors declare no competing financial interests.